\RequirePackage{ifpdf}
\ifpdf 
\documentclass[pdftex]{sigma}
\else
\documentclass{sigma}
\fi

\def\a#1{\bar{#1}}                              

\def\c#1{{\cal #1}}                             
\def\Dirac{{\raise0.09em\hbox{/}}\kern-0.69em D}

\def\epr{\epsilon}
\def\kbar{{\mathchar'26\mkern-9muk}}            
\def\lesssim{\mathrel{\hbox{\rlap{\hbox{\lower4pt\hbox{$\sim$}}}\hbox{$<$}}}}

\def\p{\partial}                                
\def\tfrac #1#2{\textstyle{\frac{#1}{#2}}}      
\def\t#1{\tilde #1}                             
\def\h#1{\hat{#1}}                              

\def\vev#1{\langle #1 \rangle}                  

\def\k{\kern-.1em\mathbin{,}\kern-.1em}
\def\hk{\kern.12em\raise-1em\hbox{$\hat{\raise1em\hbox{,}}$}\kern.12em}

\def\hI{{\hat I}}

\begin{document}

\renewcommand{\PaperNumber}{125}

\FirstPageHeading

\renewcommand{\thefootnote}{$\star$}

\ShortArticleName{WKB Approximation in Noncommutative Gravity}

\ArticleName{WKB Approximation in Noncommutative Gravity\footnote{This
paper is a contribution to the Proceedings of the Seventh
International Conference ``Symmetry in Nonlinear Mathematical
Physics'' (June 24--30, 2007, Kyiv, Ukraine). The full collection
is available at
\href{http://www.emis.de/journals/SIGMA/symmetry2007.html}{http://www.emis.de/journals/SIGMA/symmetry2007.html}}}

\Author{Maja BURI\'C~$^\dag$, John MADORE~$^\ddag$ and George ZOUPANOS~$^{\S}$}

\AuthorNameForHeading{M. Buri\'c, J. Madore and G. Zoupanos}

\Address{$^\dag$~Faculty of Physics,
               University of Belgrade, P.O. Box  368
               RS-11001 Belgrade, Serbia}
\EmailD{\href{mailto:majab@phy.bg.ac.yu}{majab@phy.bg.ac.yu}}

\Address{$^\ddag$~Laboratoire de Physique Th\'eorique,
             Universit\'e de Paris-Sud,\\
$\phantom{^\ddag}$~B\^atiment 211,
               F-91405 Orsay, France}
\EmailD{\href{mailto:madore@th.u-psud.fr}{madore@th.u-psud.fr}}

\Address{$^{\S}$~Physics Department,
            National Technical University, Zografou Campus,
              GR-15780  Athens}
\EmailD{\href{mailto:zoupanos@cern.ch}{zoupanos@cern.ch}}

\ArticleDates{Received October 25, 2007, in f\/inal form December 21, 2007; Published online December 24, 2007}

\Abstract{We consider the quasi-commutative approximation to a
  noncommutative geo\-metry def\/ined as a generalization of the moving frame formalism. The relation which \mbox{exists} between noncommutativity and geo\-metry is used to  study the properties of the high-frequency waves on the f\/lat background.}


\Classification{46L87; 83C35}

\section{Preliminary formalism}

Our purpose in this paper is to analyze the relation which exists between the noncommutativity of the local coordinates and the gravitational f\/ield on a given space-time. In  particular model of noncommutative gravity which we develop, in the noncommutative frame formalism, this relation is expressed as consistency between the algebraic and the dif\/ferential-geometric structures, or in other language as generalized Jacobi identities.

Let $\mu$ be a typical `large' source mass with `Schwarzschild radius'
$G_N\mu$. We have two length scales, determined by respectively $G_N\hbar$, the
square  of the Planck length and by $\kbar$, the scale of noncommutativity. The
gravitational f\/ield is weak if the dimensionless parameter
$\epsilon_{GF} = G_N \hbar^{-1}\mu^2$ is small; the space-time is almost
commutative if the dimensionless parameter $\epsilon = \kbar \mu^2$ is
small. These two parameters are not necessarily related but we shall
here assume that they are of the same order of magnitude,
\begin{gather*}
\epsilon_{GF} \simeq \epsilon.            
\end{gather*}
If noncommutativity is not directly related to gravity then it makes
sense to speak of ordinary gravity as the limit $\kbar \to 0$ with
$G_N\mu$ nonvanishing. On the other hand if noncommutativity and
gravity are directly related then both should vanish with $\kbar$. We
wish here to consider an expansion in the parameter $\epsilon$, which
we have seen is a measure of the relative dimension of a~typical
`space-time cell' compared with the Planck length of a typical
quantity of gravitational energy.  Our motivation for considering
noncommutative geometry as an `avatar' of gravity is the belief that
it sheds light on the role of the gravitational f\/ield as
the universal regulator of ultra-violet divergences.  We
mention here only some elements of
 the approach we use to study gravitational f\/ields on
Lorentz-signature manifolds. A
general description can be found elsewhere~\cite{Mad00c} as can a
simple explicit solution~\cite{BurMad05b}.
Some properties of the perturbation analyzed here which we shall use are derived in ~\cite{BurGraMadZou06a}.

We start with a `noncommutative space', a  $*$-algebra $\c{A}$ generated by four Hermitian elements~$x^\mu$ which satisfy the commutation relations
\begin{gather*}
[x^\mu, x^\nu] = i\kbar J^{\mu\nu}(x^\sigma) .      
\end{gather*}
We assume that algebra is associative, therefore the commutators obey the Jacobi identities.
We assume further that over $\c{A}$ there is a dif\/ferential
calculus~\cite{Mad00c}  which possesses a preferred frame~$\theta^\alpha$, a set of 1-forms which commute with the algebra,
\begin{gather*}
[x^\mu, \theta^\alpha] = 0.                            
\end{gather*}
 The space one obtains in the commutative limit has a global moving frame
$\t{\theta}^\alpha$.  The dif\/ferential is def\/ined as
\begin{gather*}
df = e_\alpha f\, \theta^\alpha,
\end{gather*}
that is,
\begin{gather*}
dx^\mu = e_\alpha^\mu \theta^\alpha, \qquad
e_\alpha^\mu = e_\alpha x^\mu .
\end{gather*}
The $e_\alpha$ are the vector f\/ields dual to the frame forms
\begin{gather*}
\theta^\alpha(e_\beta) = \delta^\alpha_\beta.
\end{gather*}

We have two basic structures: the algebra $\c{A}$
def\/ined by a product which is restricted by the matrix of elements
$J^{\mu\nu}$ and the metric  def\/ined by a frame that is the matrix of elements~$e_\alpha^\mu$.  Consistency requirements impose relations between these two structures which in
simple situations allow us to f\/ind a one-to-one correspondence between
the commutators and the metric.
Most important relation is the Leibniz rule
\begin{gather}
i\kbar d J^{\mu\nu} = [dx^\mu, x^\nu] + [x^\mu, dx^\nu] =[e^\mu_\alpha , x^\nu]\theta^\alpha - [e^\nu_\alpha , x^\mu]\theta^\alpha .   \label{lr}
\end{gather}
One can see in~(\ref{lr}) a dif\/ferential equation for $J^{\mu\nu}$ in terms of
$e^\mu_\alpha$. In important special cases this equation reduces to a
simple dif\/ferential equation of one variable.

We must insure in addition that the dif\/ferential is well def\/ined. A
necessary condition is that $\ d[x^\mu, \theta^\alpha] = 0$, from which
it follows that
\begin{gather*}
d[x^{\mu}, \theta^{\alpha}] =
[dx^{\mu}, \theta^{\alpha}] + [x^\mu, d\theta^{\alpha}] =
e^\mu_\beta [\theta^{\beta}, \theta^{\alpha}] -
\tfrac 12 [x^\mu, C^\alpha{}_{\beta\gamma}] \theta^{\beta}\theta^{\gamma}.
\end{gather*}
We  have  introduced the Ricci rotation coef\/f\/icients
\begin{gather*}
d\theta^\alpha =-\tfrac 12 C^\alpha{}_{\beta\gamma} \theta^\beta \theta^\gamma .
\end{gather*}
Therefore we f\/ind that multiplication of 1-forms  satisf\/ies
\begin{gather*}
[\theta^{\alpha}, \theta^{\beta}] =
\tfrac 12 \theta^\beta_\mu [x^\mu, C^\alpha{}_{\gamma\delta}]
\theta^{\gamma}\theta^{\delta}.                               
\end{gather*}
The metric is def\/ined by the map
\begin{gather*}
g(\theta^\alpha\otimes\theta^\beta) = g^{\alpha\beta}.
\end{gather*}
The bilinearity of the metric implies that $g^{\alpha\beta}$ are
 numbers and not functions of coordinates, exactly as in the commutative case.  We choose the frame to be orthonormal; we can write therefore
\begin{gather*}
g^{\alpha\beta} = \eta^{\alpha\beta} .  
\end{gather*}
We introduce also
\begin{gather*}
g^{\mu\nu} = g(dx^\mu \otimes dx^\nu)
= e^\mu_\alpha e^\nu_\beta g^{\alpha\beta}.
\end{gather*}
Other dif\/ferential-geometric quantities:
 connection,  torsion and  curvature can be def\/ined with the same   formulae as in  commutative dif\/ferential geometry. We will take here that the torsion vanishes.

\section{The quasi-commutative approximation}         \label{g2a}

To lowest order in $\epr$ the partial derivatives are well def\/ined and
the approximation, which we shall refer to as the quasi-commutative,
\begin{gather*}
[x^\lambda, f] =
i\kbar J^{\lambda\sigma} \p_\sigma f                
\end{gather*}
is valid.  The Leibniz rule and the Jacobi identity can be written in
this approximation as
\begin{gather*}
 e_\alpha J^{\mu\nu}
= \partial_\sigma e^{[\mu}_\alpha J^{\sigma\nu ]}  ,         
\qquad
\epsilon_{\kappa\lambda\mu\nu} J^{\gamma\lambda}e_\gamma J^{\mu\nu}=0.
\end{gather*}
We call these equations the Jacobi equations.

Written in frame components $J^{\alpha\beta}$ of the commutators, $J^{\alpha\beta} = \theta^\alpha_\mu\theta^\beta_\nu J^{\mu\nu}$,  the Jacobi equations become
\begin{gather}
e_\gamma J^{\alpha\beta}
- C^{[\alpha}{}_{\gamma\delta} J^{\beta]\delta} = 0,      \label{cl}
\\
\epsilon_{\alpha\beta\gamma\delta}J^{\gamma\eta}
(e_\eta J^{\alpha\beta}
+ C^\alpha{}_{\eta\zeta} J^{\beta\zeta}) = 0.     \nonumber         
\end{gather}
We have used here the expression for the rotation coef\/f\/icients,
valid also in the quasi-com\-mu\-ta\-ti\-ve approximation
\begin{gather*}
C^\alpha{}_{\beta\gamma}
= \theta^\alpha_\mu e_{[\beta}e^\mu_{\gamma]}
=-e^\nu_\beta e^\mu_\gamma\partial_{[\nu}\theta^\alpha_{\mu]},
\end{gather*}
and the inverse $\theta^\alpha_\mu$ of the $e^\mu_\alpha$.

Equation~(\ref{cl}) for the rotation coef\/f\/icients can be solved.  Provided $J^{-1}_{\alpha\beta}$ exists, after some algebra we obtain that
\begin{gather*}
C^\alpha{}_{\beta\gamma}
 = J^{\alpha\eta} e_\eta J^{-1}_{\beta\gamma}.            
\end{gather*}
If we introduce
\begin{gather*}
\h{C}_{\alpha\beta\gamma} =
J^{-1}_{\alpha\delta} C^\delta{}_{\beta\gamma},
\end{gather*}
we f\/ind  that
\begin{gather*}
\h{C}_{\alpha\beta\gamma} =  e_\alpha J^{-1}_{\beta\gamma}
\end{gather*}
and also
\begin{gather*}
\h{C}_{\alpha\beta\gamma} + \h{C}_{\beta\gamma\alpha}
+ \h{C}_{\gamma\alpha\beta} = 0.                
\end{gather*}
This equation we can write as a de~Rham cocycle condition
\begin{gather*}
d J^{-1} = 0, \qquad J^{-1}
= \tfrac 12 J^{-1}_{\alpha\beta} \theta^\alpha\theta^\beta. 
\end{gather*}

\section[The weak-field  approximation]{The weak-f\/ield  approximation}  \label{gf}

  We assumed in the previous section
 that the noncommutativity is small and we derived
some relations to f\/irst-order in the parameter~$\epr$.  We shall now
make an analogous assumption concerning the gravitational f\/ield; we
shall assume that $\epsilon_{GF}$ is also small and of the same order
of magnitude. With these two assumptions the Jacobi equations become
relatively easy to solve.

We suppose that the basic unknowns, the commutators and the frame components are
constants in the ground state. That is, the ground state is a f\/lat noncommutative space characterized by $J_0^{\mu\nu}$ and $\theta^\alpha_0$. It is perturbed to
\begin{gather*}
J^{\alpha\beta} = J_0^{\alpha\beta} + \epr I^{\alpha\beta}, \qquad
\theta^\alpha = \theta^\beta_0 (\delta_\beta^\alpha -
 \epr \Lambda_\beta^\alpha).
\end{gather*}
The leading order of the Jacobi system is then given by
\begin{gather}
e_\gamma I^{\alpha\beta}
- e_{[\gamma}\Lambda^{[\alpha}_{\delta]} J_0^{\beta]\delta} = 0,        \label{=clp}
\\
\epsilon_{\alpha\beta\gamma\delta}J_0^{\gamma\eta}
e_\eta I^{\alpha\beta} = 0.                                       \label{mb=p}
\end{gather}
Introducing the notation
\begin{gather*}
\h{I}_{\alpha\beta} =
J_0^{-1}{}_{\alpha\gamma} J_0^{-1}{}_{\beta\delta} I^{\gamma\delta}, \qquad
\h{\Lambda}_{\alpha\beta}
= J_0^{-1}{}_{\alpha\gamma}\Lambda^\gamma_\beta,             
\end{gather*}
(\ref{=clp}) can be written as
\begin{gather}
e_\gamma(\h{I} _{\alpha\beta} - \h{\Lambda} _{[\alpha\beta]})
= e_{[\alpha}\h{\Lambda}_{\beta]\gamma}                        \label{mb*2x}
\end{gather}
and~(\ref{mb=p})  as
\begin{gather}
\epsilon^{\alpha\beta\gamma\delta} e_\alpha \h{I}_{\beta\gamma}
= 0                  .                                           \label{c12x}
\end{gather}
We note that $\hI$ is a linear perturbation of $J_0^{-1}$,
\begin{gather*}
J_{\alpha\beta}^{-1} = J_{0\alpha\beta}^{-1} + \epr \hI_{\alpha\beta}.
\end{gather*}
Equations~(\ref{mb*2x})--(\ref{c12x}) can be solved~\cite{BurGraMadZou06a}. The most general solution is given by
\begin{gather}
\h{\Lambda}_{\alpha\beta} = \h{I}_{\alpha\beta}
+ e_\beta A_\alpha + c_{\alpha\beta},                          \label{LI}
\end{gather}
where $c$ is a
 2-form with constant components
 and $A$ is
an arbitrary 1-form.
$\hI$ must obey the cocycle condition~(\ref{mb=p}) which,
 introducing
\begin{gather*}
\h{I} = \tfrac 12 \h{I}_{\alpha\beta}\theta^\alpha \theta^\beta
\end{gather*}
can be written as
\begin{gather}
d\h{I} = 0.                  \label{dI}
\end{gather}
Therefore we see that there must exist a 1-form $C$ such that
\begin{gather*}
\h{I}_{\gamma\delta} = e_{[\gamma}C_{\delta]}.              
\end{gather*}

We can now state more precisely the relation between noncommutativity and gravity in the
linear approximation. With $c_{\alpha\beta}=0$ we have
\begin{gather}
\Lambda^\alpha_\beta
= J_0^{\alpha\gamma}( \h{I}_{\gamma\beta}
+ e_\beta A_\gamma).                                    \label{why}
\end{gather}
If we denote the perturbation of the metric as
\begin{gather*}
g^{\mu\nu} = \eta^{\mu\nu} - \epr g_1^{\mu\nu},
\end{gather*}
we easily derive the relation
\begin{gather*}
 g_1^{\mu\nu} =
- \eta^{\alpha\beta} \Lambda^{(\mu}_\alpha \delta^{\nu)}_\beta
= - \Lambda^{(\mu\nu)}.
\end{gather*}
It follows from~(\ref{why}) that
\begin{gather*}
g_1{}_{\alpha\beta} = - J_{0(\alpha}{}^{\gamma}
(\h{I}_{\gamma\beta)} + e_{\beta)} A_\gamma)       .      
\end{gather*}
The frame itself is given by
\begin{gather*}
\theta^\alpha = d ( x^\alpha -\epr J_0^{\alpha\gamma}A_\gamma)
- \epr J_0^{\alpha\gamma} \h{I}_{\gamma\beta} dx^\beta .
\end{gather*}
We therefore f\/ind  the following expressions
\begin{gather*}
 d\theta^\alpha
= - \epr J_0^{\alpha\gamma}e_\delta \h{I}_{\gamma\beta} dx^\delta dx^\beta
= \tfrac 12 \epr J_0^{\alpha\delta}e_\delta \hI_{\beta\gamma}dx^\gamma
dx^\beta ,
\qquad
C^\alpha{}_{\beta\gamma }=
\epr J_0^{\alpha\delta} e_\delta\h{I}_{\beta\gamma}.
\end{gather*}
Using the expression
\begin{gather*}
\omega_{\alpha\beta\gamma}
=\tfrac 12 (C_{\alpha\beta\gamma}
- C_{\beta\gamma\alpha} + C_{\gamma\alpha\beta})
\end{gather*}
 for the components of the connection 1-form $\ \omega^\alpha{}_\beta =\omega^\alpha{}_{\gamma\beta}\theta^\gamma$, we f\/ind
\begin{gather}
\omega_{\alpha\beta\gamma}
= \tfrac 12 \epr (J_{0[\alpha}{}^\delta e_\delta\h{I}_{\beta\gamma]}
+ J_{0\beta}{}^\delta e_\delta\h{I}_{\alpha\gamma}).  \label{oJ}
\end{gather}
The torsion obviously vanishes.

Further, using the def\/inition of the Riemann curvature tensor
\begin{gather*}
\Omega^\alpha_\beta = R^\alpha{}_{\beta\gamma\delta}\theta^\gamma\theta^\delta = d\omega^\alpha{}_\beta +\omega^\alpha{}_\gamma \omega^\gamma{}_\beta
\end{gather*}
from~(\ref{oJ}) we obtain for the linearized curvature
\begin{gather*}
R_{\alpha\beta\gamma\delta} = \tfrac{1}{2}\epr e^\eta \left(
J_0{}_{\eta[\gamma} e_{\delta]}\h{I}_{\alpha\beta}
+ J_0{}_{\eta[\alpha} e_{\beta]}\h{I}_{\gamma\delta}
\right)         .                        
\end{gather*}
For the  Ricci
curvature we f\/ind
\begin{gather}
R_{\beta\gamma } = - \tfrac{1}{2}\epr e^\zeta \left(
J_0{}_{\zeta(\beta} e^\alpha \h{I}_{\gamma)\alpha}
+ J_0^{\alpha}{}_{\zeta} e_{(\beta} \h{I}_{\gamma)\alpha} \right).
                                                        \label{vtj5}
\end{gather}
One more contraction yields the expression
\begin{gather}
R = - 2\epr J_0{}^{\zeta\alpha}e_\zeta e^\beta
\h{I}_{\alpha\beta}                                     \label{vtj6}
\end{gather}
for the Ricci scalar.  Using the cocycle condition permits us to
write this in the form
\begin{gather*}
R =  \epr  \Delta \chi ,          
\end{gather*}
where the scalar f\/ield $\chi$,  the trace component of the perturbation, is def\/ined as
\begin{gather*}
\chi = J_0^{\alpha\beta}\h{I}_{\alpha\beta}        .      
\end{gather*}

\section{The WKB approximation}                \label{wkb}

In the commutative case the WKB dispersion relations follow from the
Einstein equations. In order to introduce the WKB approximation in
noncommutative case,
we suppose that the algebra $\c{A}$ is a tensor product
\begin{gather*}
\c{A} = \c{A}_0 \otimes \c{A}_\omega
\end{gather*}
of a `slowly-varying' factor $\c{A}_0$ in which all amplitudes lie and
a `rapidly-varying' phase factor which is of order-of-magnitude $\epr$
so that only functions linear in this factor can appear. By
`slowly-varying' element $f$ of the algebra we mean an element with a classical limit $\t{f}$
such that $\p_\alpha \t{f} \lesssim \mu \t{f}$.  The generic element
$f$ of the algebra then is of the form
\begin{gather*}
f = f_0 + \epr \bar f e^{i\omega\phi},
\end{gather*}
where $f_0$ and $\bar f$ belong to $\c{A}_0$.  Because of the condition
on $\epsilon$ the factor order does not matter and these elements form
an algebra.  The frequency parameter $\omega$ is so chosen that for an
element~$f$ of~$\c{A}_0$ the estimate
\begin{gather*}
[\phi, f] \simeq \kbar \mu
\end{gather*}
holds. The commutator $[f, e^{i\omega\phi}]$ is thus of order of
magnitude
\begin{gather*}
[f, e^{i\omega\phi}] \simeq \kbar \mu\omega .
\end{gather*}
The wave vector
\begin{gather*}
\xi_\alpha = e_\alpha \phi                 
\end{gather*}
is normal to the surfaces of constant phase.  We shall require also
that the energy of the wave be such that it contribute not as source
to the background f\/ield. This inequality can be written~as
\begin{gather*}
\epsilon \omega^2 \ll \mu^2.                            
\end{gather*}
It also assures us that for the approximation we are considering we
need not pay attention to the order of the factors in the
perturbation.  We have in fact partially solved the system of
equations without further approximation. The purpose of the following
analysis is to f\/ind the constraints on the wave vector~$\xi$.

\subsection{The quasi-commutative case}

In the WKB approximation the perturbations
$\Lambda^\alpha_\beta$ and $I^{\alpha\beta}$ are of the form
\begin{gather*}
\Lambda^\alpha_\beta
= {\a{\Lambda}}^\alpha_\beta e^{i\omega\phi}, \qquad
I^{\alpha\beta} = \a{I}^{\alpha\beta} e^{i\omega\phi},
\end{gather*}
where $\a{\Lambda}^\alpha_\beta$ and $\a{I}^{\alpha\beta}$ belong to
$\c{A}_0$.  Therefore we have also
\begin{gather*}
g_1^{\mu\nu} = \a{g}^{\mu\nu}e^{i\omega\phi} .
\end{gather*}
Using $\xi_\alpha$ and $
\eta^\alpha = J_0^{\alpha\beta} \xi_\beta    $ we have
\begin{gather*}
e_{\alpha} I_{\beta\gamma}
= (i\omega\xi_\alpha \a{I}_{\beta\gamma}
+ e_{\alpha}\a{I}_{\beta\gamma}) e^{i\omega\phi} ,
\qquad
e_{\alpha} \Lambda_{\beta\gamma}
= (i\omega\xi_\alpha \a{\Lambda}_{\beta\gamma}
+ e_{\alpha}\a{\Lambda}_{\beta\gamma}) e^{i\omega\phi}.
\end{gather*}

The cocycle condition replaces Einstein equation to a certain
extent. In the WKB approximation it becomes
\begin{gather*}
\xi_\alpha \h{I}_{\beta\gamma}
+ \xi_\beta \h{I}_{\gamma\alpha}
+ \xi_\gamma \h{I}_{\alpha\beta} = 0.                    
\end{gather*}
We multiply this equation by $\xi^\alpha$ and obtain
\begin{gather}
\xi^2 \h{I}_{\beta\gamma}
+ \xi_{[\beta} \h{I}_{\gamma]\alpha}\xi^\alpha = 0.      \label{cyc2}
\end{gather}
If $\xi^2 \neq 0$ then we conclude that
\begin{gather*}
\h{I}_{\beta\gamma}
= - \xi^{-2}\xi_{[\beta} \h{I}_{\gamma]\alpha}\xi^\alpha.
\end{gather*}
This is no restriction; it def\/ines simply $C_\alpha$ by
\begin{gather*}
i\omega C_{\alpha} = - \xi^{-2} \h{I}_{\alpha\beta}\xi^\beta.
\end{gather*}
If $\xi^2 = 0$ then we conclude that
\begin{gather*}
\xi_{[\beta} \h{I}_{\gamma]\alpha}\xi^\alpha = 0.
\end{gather*}
This is a small restriction; the $\xi_\alpha$ must be a Petrov vector
of $\h{I}$. We shall improve this in a~particular case in the next
section.  In terms of the scalar $\chi$ we obtain the
relation
\begin{gather}
\h{I}_{\alpha\beta}\eta^\beta
= -\tfrac 12 \chi \xi_\alpha.                           \label{ec}
\end{gather}

Using the def\/inition of $\eta$ we f\/ind in the WKB approximation to
f\/irst order
\begin{gather}
\omega_{\alpha\beta\gamma}
= \tfrac 12 \epr (i\omega)\left(\eta_{[\alpha} \h{I}_{\beta\gamma]}
+\eta_{\beta} \h{I}_{\alpha\gamma}\right),
\nonumber\\
R_{\alpha\beta\gamma\delta} =- \tfrac 12\epr (i\omega)^2
\left(\eta_{[\gamma}\xi_{\delta]} \hI_{\alpha\beta}
- \eta_{[\alpha}\xi_{\beta]} \h{I}_{\gamma\delta} \right) ,
\label{R}\\
R_{\beta\gamma} =- \tfrac 12\epr (i\omega)^2
\left(\xi_{(\beta} \eta^\alpha
- \xi^{\alpha} \eta_{(\beta}\right) \h{I}_{\gamma)\alpha} ,
\nonumber\\
 R =  \epr(i\omega)^2\chi\xi^2 .        \nonumber      
\end{gather}
In average the terms linear in $\epr$ vanish. Therefore in principle we have to calculate to second order and average over several wavelengths. Using
\begin{gather*}
\vev{\h{I}^{\alpha\beta}} = 0, \qquad
 \vev{\h{I}^{\alpha\beta} \h{I}^{\gamma\delta}}
= \tfrac 12 \h{\a{I}}^{\alpha\beta} \h{\a{I}}^{\gamma\delta} ,
\end{gather*}
and expanding the curvature to second order we f\/ind
the expression
\begin{gather}
\vev{R_{\beta\gamma}}
= \tfrac 12\epr^2(i\omega)^2
\Big( \a{\chi} \xi^\alpha\eta_{(\gamma }\h{\a{I}}_{\beta )\alpha}
+ \tfrac 34  \a{\chi}^2 \xi_\beta\xi_\gamma
+\eta^2\h{\a{I}}_{\eta\beta}\h{\a{I}}^\eta{}_\gamma
-\tfrac 12 \eta_\beta \eta_\gamma
\h{\a{I}}_{\alpha\eta}\h{\a{I}}^{\alpha\eta} \Big) \label{Ric}
\end{gather}
for the Ricci tensor and the expression
\begin{gather*}
\vev{R} = \tfrac 18 \epr^2 (i\omega)^2
(2 \eta^2 \h{\a{I}}_{\alpha\beta}\h{\a{I}}^{\alpha\beta}
+ 7\a{\chi}^2\xi^2)
\end{gather*}
for the Ricci scalar. We shall return to these formulae in
Section~\ref{vtjam}.

\subsection{The noncommutative lattice}                    \label{ncl}

As a lattice, the background noncommutativity is of considerable
complexity, the contrary of a~simple cubic lattice. It is in general
non-periodic but in the WKB approximation we can assume periodicity
since at the scale of the frequency $J_0^{\mu\nu}$ is a constant $4\times4$
matrix. It is dif\/f\/icult to obtain general expressions for the modes of the high frequency waves and
their dispersion relations; however, it is interesting to analyse them in more
detail by considering a~specif\/ic example. We take an arbitrary
perturbation $\hI_{\alpha\beta}$ with the wave vector $\xi_\alpha$
normalized so that $\xi_0= -1$,
\begin{gather*}
\h{\a{I}}_{\alpha\beta} =
\left(\begin{array}{cccc}
   0  &  b_3 & -b_2 & e_1 \\[2pt]
 -b_3 &  0   &  b_1 & e_2 \\[2pt]
  b_2 & -b_1 &  0   & e_3 \\[2pt]
 -e_1 & -e_2 & -e_3 &  0
\end{array}\right) .
\end{gather*}
One easily sees that the cocycle condition is equivalent to the
constraint $\vec b = - \vec\xi \times \vec e$ which is the part of the
f\/ield equations for the electromagnetic plane wave, the Bianchi equations.
Suppose that $\xi$ is null and oriented along the $z$-axis,
$\xi_\alpha = (0,0,1,-1)$. The cocycle condition imposes that $\hI$ equals to
\begin{gather}
\h{\a{I}}_{\alpha\beta} =
\left(\begin{array}{cccc}
   0  &  0   & -e_1 & e_1 \\[2pt]
   0  &  0   & -e_2 & e_2 \\[2pt]
  e_1 & e_2  &  0   & e_3 \\[2pt]
 -e_1 & -e_2 & -e_3 &  0
\end{array}\right)  .                            \label{matI}
\end{gather}
The perturbation $\hI$ is of Petrov-type $N$ if $\vec\xi\cdot\vec e=e_3=0\,$;
this would be the second half of the Maxwell f\/ield equations.  In
this case, for an arbitrary background noncommutativity given by
\begin{gather}
J_{0\alpha\beta}=
\left(\begin{array}{cccc}
  0  &  B_3 & -B_2 & E_1 \\[2pt]
-B_3 &  0   &  B_1 & E_2 \\[2pt]
 B_2 & -B_1 &  0   & E_3 \\[2pt]
 - E_1 &  -E_2 &  -E_3 &  0
\end{array}\right)                               \label{matJ}
\end{gather}
we can write the amplitude of the metric perturbation in the form
\begin{gather*}
\a{g}_{1\alpha\beta}
= - J_{0(\alpha}{}^{\gamma}\h{\a{I}}_{\gamma\beta)} =
\left(\begin{array}{cc}
P_{11}&P_{12}\\[2pt] P^T_{12}&P_{22}
\end{array}\right).
\end{gather*}
The expressions for $P_{12}$, $P_{22}$ can be obtained from~(\ref{matI})--(\ref{matJ}) but they are somewhat lengthy. However,
it is easy to check that by a change of coordinates we can set
$P_{12} =0$, $P_{22} =0$. Introducing
$e_1 = a\cos \gamma$, $e_2 = a\sin\gamma$,
$B_2+E_1 = A\sin \Gamma$, $B_1-E_2 = A\cos\Gamma$
the remaining part~$P_{11}$
 can be decomposed
\begin{gather*}
 P_{11} =
aA\left(\begin{array}{cc}
\sin(\gamma+\Gamma)    &\cos(\gamma+\Gamma)\\[2pt]
\cos(\gamma+\Gamma)    &-\sin(\gamma+\Gamma)
\end{array}\right)
+
aA\left(\begin{array}{cc}
\sin(\gamma-\Gamma)    &         0\\[2pt]
         0             &\sin(\gamma-\Gamma)
\end{array}\right)
\end{gather*}
into a trace-free part and a trace. The
trace-free part corresponds to a
 gravitational wave which is polarized, and though the
polarization is f\/ixed in terms of $\gamma+\Gamma$,
it can be arbitrary.
In addition there is a scalar wave, the trace. In the case when $e_3\neq 0$ the perturbation $\h{I}$ is not of Petrov
type~$N$; the additional gravitational mode is longitudinal, a constraint mode.

\section{The Poisson energy and conservation laws}                    \label{vtjam}

We have in fact associated a gravitational f\/ield to the noncommutative
structure with the map~(\ref{LI}). We would like to consider now this
structure as an ef\/fective f\/ield and estimate its
energy-momentum which we call the `Poisson energy'.  We are confronted immediately with the choice of the position of the extra term
in the Einstein equations. If we place it on the right-hand side, we
can consider it  as an ef\/fective matter source. If we
 keep it on the left-hand side then we interpret it as a noncommutative
modif\/ication of the curvature. First however we make some preliminary remarks about conservation laws.

From (\ref{vtj5})--(\ref{vtj6}) for the Einstein tensor we obtain
\begin{gather*}
G_{\beta\gamma}
= -\tfrac 12\epr\Big(J_{0\zeta(\beta}
e^\zeta e^\alpha \h{I}_{\gamma)\alpha}
+ J_0{}^{\alpha\zeta}e_\zeta e_{(\beta}\h{I}_{\gamma)\alpha}
- 2\eta_{\beta\gamma}J_{0\zeta\delta}
e^\zeta e_\alpha\h{I}^{\delta\alpha}\Big).              
\end{gather*}
In general the Einstein tensor does not vanish.  A conservation equation of the associated energy-momentum tensor in  linear  approximation is easy to verify.  Applying the cocycle
condition and keeping in mind that, to linear order in $\epr$,
$e_\alpha e_\beta = e_\beta e_\alpha$, we obtain
\begin{gather*}
e^\beta G_{\beta\gamma} = -\tfrac 12\epr \Big(
J_{0}{}^{\alpha\zeta}e^\beta e_\zeta e_\gamma \h{I}_{\beta\alpha}
+J_{0}{}^{\alpha\zeta}e^\beta e_\zeta e_\beta \h{I}_{\gamma\alpha}
-2J_{0\zeta\delta}e_\gamma e^\zeta e_\alpha\h{I}^{\delta\alpha}\Big)
\nonumber \\
\phantom{e^\beta G_{\beta\gamma}}
=- \tfrac 12\epr J_0{}^{\delta\zeta}
e_\zeta e^\alpha ( e_\alpha \h{I}_{\gamma\delta}
-e_\gamma \h{I}_{\alpha\delta})
= \tfrac 12\epr J_0{}_{\delta\zeta}e^\zeta e^\alpha e^\delta
\h{I}_{\alpha\gamma} = 0.
\end{gather*}
As we shall see, the conservation law holds in an important special
case in quadratic order too.

\subsection{Canonical orientation}

To the extent that the noncommutative background is analogous to a
lattice, the perturbations can be considered as elastic vibrations or
phonons. This analogy however is tenuous at the approximation we are
considering since we have excluded any resonance phenomena. These
could appear if we allowed larger-amplitude waves with energy
suf\/f\/icient to change the background. The case we shall now focus to
would then be analogous to a phonon propagating along one of the axes of a
regular cubic lattice. In the special case in which it is
also Petrov vector of the perturbation the dispersion relations become
clearer.

Assume then that $\eta$ and $\xi$
are parallel and set
\begin{gather}
\eta^\alpha = J_0^{\alpha\beta}\xi_\beta =\lambda \xi^\alpha.                       \label{cash}
\end{gather}
It follows from~(\ref{ec})  that the vector $\xi$ is an
eigenvector of $J$ not only to f\/irst but also to second order.
Equation~(\ref{R}) yields for the linearized Riemann curvature
\begin{gather*}
R_{\alpha\beta\gamma\delta} =0.
\end{gather*}
The dispersion relation
\begin{gather*}
\xi^2 = 0
\end{gather*}
follows from~(\ref{cyc2}).

In quadratic order, using the dispersion relation, we f\/ind that the
expression~(\ref{Ric}) for the Ricci tensor simplif\/ies to
\begin{gather*}
\vev{R_{\beta\gamma}} = - \tfrac 18\epr^2(i\omega)^2
(\a{\chi}^2 + 2\lambda^2 \h{\a{I}}_{\alpha\eta}\h{\a{I}}^{\alpha\eta})
\xi_\beta\xi_\gamma.
\end{gather*}
 The  Ricci
scalar vanishes and we obtain for the Einstein tensor the average value
\begin{gather*}
  \vev{G_{\beta\gamma}} = -\rho \xi_\beta\xi_\gamma     
\end{gather*}
with
\begin{gather*}
\rho = -\tfrac 18(\epr\omega)^2
(\a{\chi}^2 + 2\lambda^2 \h{\a{I}}_{\alpha\eta}\h{\a{I}}^{\alpha\eta}) .
\end{gather*}
The energy-momentum is that of a null dust with a density $\rho$.

In the WKB approximation we can, just as in the classical case, derive
a conservation law for $\rho$ which has a natural interpretation as
graviton-number conservation.  If we multiply the
cocycle condition~(\ref{dI}) by $\xi^\alpha$ we obtain
\begin{gather*}
\xi^\alpha e_\alpha \h{I}_{\beta\gamma}
+ \xi^\alpha e_\beta \h{I}_{\gamma\alpha}
+ \xi^\alpha e_\gamma \h{I}_{\alpha\beta} = 0.       
\end{gather*}
We also have
\begin{gather*}
 e^\alpha(\xi_\alpha \h{I}_{\beta\gamma}
+ \xi_\beta \h{I}_{\gamma\alpha}
+ \xi_\gamma \h{I}_{\alpha\beta}) = 0.   
\end{gather*}
Adding these two equations, using~(\ref{ec}), (\ref{cash}) and not
forgetting that
$e_\alpha\xi_\beta = e_\beta\xi_\alpha$ in our approxi\-mation,
we f\/ind
\begin{gather*}
 (\xi^\alpha e_\alpha \h{I}_{\beta\gamma}
+ e^\alpha( \xi_\alpha \h{I}_{\beta\gamma})) \h{I}^{\beta\gamma}
+2 (\xi^\alpha e_\beta \h{I}_{\gamma\alpha}
+  e^\alpha( \xi_\beta \h{I}_{\gamma\alpha})) \h{I}^{\beta\gamma}
\nonumber\\
\qquad{}=e_\alpha(\xi^\alpha\hI^{\beta\gamma }\hI_{\beta\gamma})
+ 2e^\alpha(\xi^\beta\hI^{\gamma\alpha}\hI_{\beta\gamma}) =0  .
\end{gather*}
The conservation law
\begin{gather*}
e^\alpha (\rho \xi_\alpha) = 0
\end{gather*}
follows and from it the conservation of the ef\/fective source,
\begin{gather*}
e^\alpha(\rho \xi_\alpha\xi_\beta) =0.
\end{gather*}

To interpret the additional term we have isolated
as the energy-momentum of an external f\/ield,
\begin{gather*}
G_{\beta\gamma} = -16\pi G_N  T_{\beta\gamma}
\end{gather*}
the sign of $\rho$ should be non-negative.  However, as
\begin{gather*}
\rho = -\tfrac 18(\epr\omega)^2
(\a{\chi}^2 + 2\lambda^2 \h{\a{I}}_{\alpha\eta}\h{\a{I}}^{\alpha\eta})
= \tfrac 14(\epr\omega\lambda{\vec e})^2
- \tfrac 14(\epr\omega\lambda{\vec b})^2
 - \tfrac 18(\epr\omega\a{\chi})^2  ,
\end{gather*}
the matter density does not have a f\/ixed sign, unless of course one
place restrictions on the relative importance of the space-time and
space-space commutation relations.
This exactly is one of the properties which could explain the
acceleration of the universe~\cite{SahSta06} and it makes the `Poisson energy'
a possible candidate for dark energy. We shall examine
this in more detail in the future work.

\section{Conclusions}

The formalism on which the article has been based is one with a
preferred frame. It is in a sense gauge-f\/ixed from the
beginning. We have shown that the degrees-of-freedom or basic modes of
the resulting theory of gravity can be put in  correspondence
with those of the noncommutative structure. As an application of the
formalism we have considered a high-frequency perturbation of the
metric.  In the classical theory it follows from the f\/ield equations
that the perturbation must satisfy a dispersion relation and a
conservation law. We show that these remain valid in the
noncommutative extension of the frame formalism and that they are
consequences of a~cocycle condition on
the corresponding perturbation of the Poisson structure.

The analysis of the dispersion relations however shows that the content of the high-frequency radiation is not identical in the two cases. Noncommutative gravity accommodates the Einsteinean gravitational waves but they get necessarily polarized by the background noncommutative lattice.  There is also a massless scalar mode. In addition, we obtain that massive longitudinal modes can exist; it is reasonable however to expect that they would be eliminated by some additional equations of motion for the Poisson structure.

We have also shown that the perturbation of the Poisson structure
contributes to the energy-momentum as an additional ef\/fective source
of the gravitational f\/ield. Although the explicit form of this
contribution, the Poisson energy, was calculated only in a linearized,
high-frequency approximation it is certainly
signif\/icant in a more general context. It would be very interesting to examine the properties the Poisson energy beyond the WKB approximation, in particular in the context of cosmology.

It would be nice, as a referee of this paper has suggested, to compare the details of the frame approach to noncommutative gravity to the other approaches, e.g.~\cite{Ca,Ga,Asc1,Asc2}; we will here point out the main dif\/ferences. A careful reader has already noticed that the noncommutative frame formalism has intrinsically geometric formulation. The metric, connection, curvature are all def\/ined via the forms; they have the usual dif\/ferential-geometric properties as e.g.\ linearity, derivatives obey the Leibniz rule etc. As a consequence, inbuilt in the theory are the usual symmetries as coordinate invariance; the commutator $J^{\mu\nu}$ can have arbitrary dependence on the coordinates. On the other hand, the analysis as presented is representation-free and thus we have no action or equations of motion: all relations are just algebraic constraints.

The approach developed in \cite{Ca,Ga,Asc1,Asc2} emphasizes  the gauge-f\/ield properties of the vielbein and connection and follows the logic of the f\/ield theory. Thus basically one needs the representation of the noncommutative f\/ields (usually, the Moyal--Weyl representation with the commutator $J^{\mu\nu}$= const). Some properties of the dif\/ferential calculus have to be changed, for example the Leibniz rule. Symmetries however are well def\/ined and the action principle can be postulated. The two approaches, clearly, dif\/fer conceptually and maybe the best way to compare them is to compare the corresponding solutions to the specif\/ic physical problems.
In this spirit, we present our
 results for the gravitational wave propagation here.

\subsection*{Acknowledgements}

This work is supported by the EPEAEK programme ``Pythagoras II'' and
co-funded by the European Union(75\%) and the Hellenic state (25\%). A CEI grant for participation in the Seventh International Conference ``Symmetry in Nonlinear Mathematical Physics''  is  gratefully acknowledged.

\pdfbookmark[1]{References}{ref}
 \LastPageEnding
\end{document}